\documentclass[12pt]{article}
\usepackage{fullpage}
\usepackage{graphicx}
\usepackage{amsmath}
\usepackage{listings}
\newcommand{\be}{\begin{equation}}
\newcommand{\ee}{\end{equation}}

\begin{document}
\title{Changing the paradigm of fixed significance levels: Testing Hypothesis by Minimizing Sum of Errors Type I and Type II}
\author{Pericchi, Luis(1) and Pereira, Carlos(2) \\ {\small
 1) Department of Mathematics and Center for Biostatistics and
Bioinformatics,} \\ {\small University of Puerto Rico, Rio Piedras,
San Juan, PR} \\ {\small 2) Instituto de Matematica e Estatistica,
Univ. Sao Paulo, Brasil}}
\date{May 30, 2013}
\maketitle

\begin{abstract}
Our purpose, is to put forward a change in the paradigm of testing by generalizing a very natural idea exposed by Morris DeGroot (1975) aiming to an approach that is attractive to all schools of statistics, in a procedure better suited for the needs of science. DeGroot's seminal idea is to base testing statistical hypothesis on minimizing the weighted sum of type I plus type II error instead of of the prevailing paradigm which is fixing type I error and minimizing type II error. DeGroot's result is that in simple vs simple hypothesis the optimal criterion is to reject, according to the likelihood ratio as the evidence (ordering) statistics using a \textbf{fixed} threshold value, instead of a \textbf{fixed} tail probability. By defining expected type I and type II errors, we generalize DeGroot's approach and find that the optimal region is defined by the ratio of evidences, that is, averaged likelihoods (with respect to a prior measure) and a threshold fixed. This approach yields an optimal theory in complete generality, which the Classical Theory of Testing does not. This can be seen as a Bayes-Non-Bayes compromise: the criteria (weighted sum of type I and type II errors) is Frequentist, but the test criterion is the ratio of marginalized likelihood, which is Bayesian. We give arguments, to push the theory still further, so that the weighting measures (priors)of the likelihoods does not have to be proper and highly informative, but just predictively matched, that is that predictively matched priors, give rise to the same evidence (marginal likelihoods) using minimal (smallest) training samples.\\
The theory that emerges, similar to the theories based on Objective Bayes approaches, is a powerful response to criticisms of the prevailing approach of hypothesis testing, see for example Ioannidis (2005) and Siegfried (2010) among many others.
\end{abstract}

\section{Changing the Paradigm of Hypothesis Testing and Re-visiting Bayes Factors and Likelihood Ratios}
\subsection{Introduction}

Classical significance testing, as developed by Neyman and Pearson, is suited and was designed to perform very specific comparisons, under well designed studies, on which beforehand, based on a specified Type I error (false rejection), which is restricted to be $\alpha$, and a most powerful statistics is found so that the Type II error $\beta$ is minimized. The sample sizes are chosen so that $\beta$ is bigger than or at least of the same order than $\alpha$. But the vast majority of studies do not conform to this standard. Or even if individual studies conform to the standard, merging studies no longer do. Fixing Type I error, for whatever amount of evidence and fixed tables of p-values are not justifiable (at least when there is an explicit or implicit alternative hypothesis), since then the Type II error is completely outside of control, with the possibility that Type I error may be enormous as compared with Type II error.\\
There is the need for an alternative paradigm to: i) Fix Type I error at $\alpha$ and Minimize Type II error, or ii) Calculate p-value and interpret it as the minimum $\alpha$ for which you will reject the null hypothesis, using a fixed table of values like 0.1*, 0.05**, 0.01***.

Morris DeGroot in his authoritative book (1975), Probability and Statistics 2nd Edition, perhaps the best bridge between schools of statistics ever written, stated that it is more reasonable to minimize a weighted sum of Type I and Type II error than to specify a value of type I error and then minimize Type II error. He showed it beyond reasonable doubt, but only in the very restrictive scenario of simple VS simple hypothesis. We propose here a very natural generalization for composite hypothesis, by using general weight functions in the parameter space. This was also the position taken by Pereira (1985). We show, in a parallel manner to DeGroot's proof and Pereira's discussion, that the optimal test statistics are Bayes Factors, when the weighting functions are priors with mass on the whole parameter space and loss functions which are constant under each hypothesis. When there are areas of indifference (namely areas of no practical importance, like "the new drug is interesting only if it is at least $10\%$, say, more effective than the gold standard"), then loss functions that are equal to zero in the indifference region (i.e. $0$ to $10\%$) achieve the goal of practical instead of statistical significance.\\
Hypothesis testing is  by far the more contentious aspect of statistics. There is no agreement between the schools of statistics, and neither within the Bayesians or Frequentists. It is time to shift to DeGroot's paradigm, to meet the needs of science. We show here that the alternative theory yields general optimal test, compared with the traditional theory on which the existence of optimal test is rather the exception than the rule.

We present only very simple examples, for the sake of clarity of a general argument.
\subsection{Why significance test works for carefully designed studies but not otherwise?}
Designed studies for testing hypotheses following classical significance testing are based on a careful balance between
controlling Type I Error and Minimizing Type II error. \\
Consider the following example motivated by DeGroot (1975),Section 8.2. \\
\textbf{Example 1:}Suppose we have Normal data with scale $\sigma_0=3$, and we are interested in comparing two possible means:
\[
H_0: \theta=\theta_0=-1 \mbox{  VS  } H_1:\theta=\theta_1=1.
\]
It is desired to design an experiment (the observations are costly) so that Type I error (False Rejection of $H_0$)is $0.05$ and Type II error
(False acceptance of $H_0$)is $0.1$.  Application of the Classical Neyman-Pearson Lemma yields, that the optimal criteria is based on the likelihood ratio:
\[
\mbox{ Reject } H_0 \mbox{ if }: \frac{\bar{x}-(-1)}{\sigma_0/\sqrt{n}} \ge C_{\alpha}.
\]
 Now the constant $C_{\alpha}$ is chosen as to have Type I Error equal $0.05$, that is:
 \[
 Pr(\frac{\bar{X}-(-1)}{\sigma_0/\sqrt{n}} \ge C_{\alpha}|H_0)=\alpha,
 \]
  which is immediately recognized as the familiar
 $C_{\alpha=0.05}=1.645$. Thus going to the Type II requirement then
 \[
 \beta=Pr(\frac{\bar{X}-(-1)}{\sigma_0/\sqrt{n}}<1.645|H_1)=0.1,
 \]
 gives $n = 19.25$ so we settle for $n=20$, as our designed experiment, resulting in $\beta=0.091$. This implies that $H_0$ is rejected if $\bar{X}>0.1$. Notice that in this situation $\alpha/\beta=0.55$, letting about a $1$ to $2$ relative sizes of Type I over Type II error.
 \subsection{The conundrum of "an approach bothered by good information"} Example 1 (continued): after you gave the researchers your design, they come back to you, and proudly give you $N=100$ data, since it costed the same to produce $n=20$ than $N=100$ (a situation which is not that unusual). The researchers are very satisfied with their prolific experiment, but the statistician is disturbed. As usual Type I error is kept fixed and equal to $\alpha=0.05$, but... What is the new Type II error? The statistician makes a calculation and obtains that the new type II error is $\beta=0.00000026$, or
  $\alpha/\beta=195217$, quite different from $1$ to $2$ as designed. In fact the rejection region became: $\bar{x}> -0.51$. Thus if for example the observed $\bar{x}= -0.5$ the Null Hypothesis $H_0: \theta=-1$ is rejected in favor of an alternative much further away to the observed value. This opens a conundrum: why more information is a bad thing for the traditional approach of hypothesis testing?. Incidentally, in the perhaps more frequent situation on which information is lost, for example due to drop-outs, for example the real sample size was really $n=10$, then type II error is inflated to $\beta=0.32$ if $\alpha$ is kept fixed, unbalancing in the opposite direction type I and type II errors.\\

  Example 1 illustrate why significance testing is inadequate for measuring the evidence in favor or against a hypothesis in general. It is timely to go back to the essentials. \\
  Morris DeGroot (1975), argued that instead of fixing Type I error (or computing a p-value with an scale held fixed) and minimize Type II, a better hypothesis testing paradigm is to choose the test as to minimize a weighted sum of the errors, namely
  \be
  Min_{\delta} \;\;\;[ a \cdot \alpha_{\theta_0}(\delta) + b \cdot \beta_{\theta_1}(\delta) ],
  \ee
  where $\delta$ denotes the test: ${\delta(\textbf{x})=I_{\texttt{R}}}(\textbf{x})$, where ${\texttt{R}}$ is the Rejection Region of $H_0$, $I_S(y)$ is the indicator function, equal to $1$ if and only if $y\in S$. Notice the apparently slight difference of (1) with the traditional approach of Significance Testing of fixed significance level $\alpha_0$:
  \be
  \mbox{ Restricting to those tests  }\delta \mbox{ on which Type I error: } \alpha_{\theta_0}(\delta) \le \alpha_0, \mbox{ Min}_{\delta} \;\; \beta_{\theta_1}(\delta).
  \ee
  However, the difference is far reaching as we will see in the sequel.
  In the following the superiority of DeGroot's approach is distinctly apparent. \\
  \textbf{Example 1 (continued):} In the simple vs simple  hypothesis DeGroot's proves that the optimal test is:
  \[
  \mbox{ Reject $H_0$ if } \frac{f(\textbf{x}|\theta_0)}{f(\textbf{x}|\theta_1)} < \frac{b}{a}.
  \]
  Now this approach can handle effectively \textbf{any} sample size, as long of course, that we are prepared to select $a$ and $b$. This has been perhaps the most important contention, not to embrace a more balanced and sensible combination of the two types of error. The point we make here is that the choice of $b/a$ has been already done! To see this lets go back to the design situation on which the sample size was chosen to be $n=20$. Now for $\alpha = 0.05$ and $\beta=0.091$, DeGroot's rejection region is equivalent to, after some algebra, to
  \be
  \mbox{ Reject if    } \exp(\frac{n}{\sigma^2_0} (\theta_1-\theta_0)[(\theta_0+\theta_1)/2 - \bar{x}]) < \frac{b}{a},
  \ee
  and since the rejection region is $\bar{x}>0.1034$, that is equivalent in our particular example to  $-\frac{3}{2\sqrt{20}}\cdot \log(\frac{b}{a})+\sqrt{20}/3=1.645$, obtaining $\frac{b}{a}=0.63$. Thus if we set $a=1$, the implicit value of $b$ is $0.63$. Now DeGroot's approach leads to a criterion that makes sense for any sample size, $n$
  \[
  \mbox{The Optimal Rejection Region \texttt{R} is:   } \bar{x}>\frac{9}{2 \cdot n}\times 0.46,
  \]
  with cutting point that is always positive, but approaching zero, as it is intuitively reasonable for $\theta_0=-1$ and $\theta_1=1$, and $a>b$.
  Furthermore now the ratio between
  $\alpha=1-\Phi(3\cdot.23/\sqrt(n)+\sqrt{n}/3)$ and $\beta=\Phi(3\cdot.23/\sqrt(n)-\sqrt{n}/3)$ as a function of $n$, is extremely stable ranging from $0.55$ at $n=20$ to $0.61$ at $n=100$. Thus we have found that an $\alpha$ of $0.05$ for $n=20$, is equivalent to an $\alpha$ of $0.00033$
  for $n=100$. This shows how unbalanced is the usual method of letting $\alpha$ unchanged whatever the information. Notice that not even changing to $\alpha=0.01$ would have been an effective remedy for $n=100$, since the equivalent $\alpha$ is about thirty times lower.\\
  NOTE: The previous analysis opens a interesting method to ``decrease $\alpha$ with $n$". Notice that from the formula of $\alpha$ above, using Mill's ratio, we get the following simple approximation:
  \be
  \alpha_n \sim 1-\Phi(\sqrt{n}/\sigma_0) \approx \frac{\phi(\sqrt{n}/\sigma_0)}{\sqrt{n}/\sigma_0},
  \ee
  giving a clear guidance on how to decrease the scale of p-values with the sample size. Notice how fast the p-values ought to decrease with the sample size to give a comparable amount of surprise against the model. In Section 7 we will see that the rate of decrease is different (much slower) for more complex tests.

  \subsection{The Lindley Paradox is not necessarely a difference between Bayes and Non-Bayes but between fixed significance levels and minimizing the weighted sum or errors}
Lindley's Paradox, Lindley (1957)
has been understood as the increasing divergence (as the information accumulates) between Classical Hypothesis Testing and Bayesian Testing evidence measures. There is also a divergence between DeGroot's and Classical testing, even when there are no prior densities.\\
To se this, we go back to the motivating example, of simple hypothesis against a simple alternative,  as the simplest setting on which it emerges clearest that the discrepancy is due to the criterion to be minimized. If you relinquish fixed significance levels, and adopt DeGroot's approach of minimizing
the weighted sum of errors then there is a $1 \mbox{ to } 1$ relationship with Bayesian posterior model probabilities (as it is with testing based on the Likelihood Ratio).\\
To see this recall that according to minimizing the weighted sum of errors approach the criterion is (1).\\
\textbf{Example 1 (continued):}
We have in Example 1, that the optimal rejection region, according to DeGroot's criterion can be written as
\be
\mbox{ Reject if: } \frac{\bar{x}+1}{\sigma_0/\sqrt{n}}\ge \frac{2.07}{\sigma_0 \sqrt{n}} + \frac{\sqrt{n}}{\sigma_0},
\ee
to be compared to the traditional rejection rule with fixed significance levels
\be
\mbox{ Reject if: } \frac{\bar{x}+1}{\sigma_0/\sqrt{n}}\ge 1.645.
\ee
This is the divergence or Lindley's Paradox but among two frequentist rejection rules. It is to be noted the striking difference in behavior of the two hand right sides: the fact is that under DeGroot's criterion as n grows the type I error goes to zero (as does the Type II error, and so consistency is achieved), but in the traditional approach the type I error is kept fixed and consistency fails (there is a positive probability of the wrong decision, false rejection no matter how large n is).\\
On the other hand, a general $1$ to $1$ relationship can be established between minimizing the weighted sum of errors and the probability of the null hypothesis. For any simple VS simple comparison, if $\pi_0$ and $\pi_1$ are respectively
the prior probabilities of the Null and the Alternative, then Bayes Theorem yields as the posterior probability of the Null:
\be
P(H_0|x)=\frac{\pi_0 f(x|\theta_0)}{\pi_0 f(x|\theta_0)+ \pi_1 f(x|\theta_1)}=[1+\frac{\pi_1 f(x|\theta_1)}{\pi_0 f(x|\theta_0)}]^{-1}.
\ee
Therefore if the ratio $b/a$ is interpreted as $\pi_1/\pi_0$, then the DeGroot's rejection region is equivalent
with the region on which rejection of the null occurs if $P(H_0|x)<0.5$. No divergence here, on the contrary a perfect correspondence.

 \subsection{A More General Setting}
Suppose we are testing the following two general hypotheses:
\be
H_0: \theta\in\Theta_0
\mbox{ VS } H_1: \theta\in\Theta_1
\ee
We define, in the Neyman-Pearson tradition, Type I and Type II error, of the test $\delta$ at the parameter point $\theta$ as,
\be
\alpha_\theta(\delta)=Pr(\mbox{Rejecting }H_0|\theta\in\Theta_0),
\ee
\be
\beta_\theta(\delta)=Pr(\mbox{Accepting }H_0|\theta\in\Theta_1).
\ee
\textbf{Definition}: The weighted (or expected) Type I and Type II errors are defined respectively as:
\be
\alpha(\delta) = \int_{\Theta_0}\alpha_\theta(\delta) \pi_0(\theta) d\theta,
\ee
and
\be
\beta(\delta)=\int_{\Theta_1}\beta_\theta(\delta) \pi_1(\theta) d\theta,
\ee
where $\pi_j(\theta)\geq 0$ are such that $\int_{\Theta_j}\pi_j(\theta)=1, j=0,1$. Why the expectation?: It is important to note that the errors depends on $\theta$, which is obviously unknown. How to deal with it?
In Berger (2008) it is stated that: "There are many situations in which it has been argued that a frequentist (statistician) should use an average
coverage criterion; see Bayarri and Berger (2004) for examples and references." Similarly we argue here that both Bayesians and Frequentist, should average errors. Among the main reasons we have: i) averaging errors permits a completely general theory of optimality (as we will see), ii) averaging is natural (from a probabilistic point of view, it is the marginal error) and it is flexible, in the choice of weight functions iii) the methodologies for assessing weighings have advanced, see for example Pereira (1985), Berger and Pericchi (1996, 2001), iv) it is a natural mix of Frequentist errors and Bayesian averaging.
\subsubsection{Interpretations of the weight (prior) measures}
We pause here to discuss interpretations, since there are various possible interpretations of the weight measures $\pi_j(\theta), j=0,1$.
\begin{enumerate}
\item{Prior Measures:} The most obvious is the assumed prior density of the parameter values conditional on each hypothesis, which is the natural interpretation under a Bayesian framework. Notice that not necessarily this interpretation leads to a subjective approach. If a general method for generating conventional priors like the Intrinsic prior method, then this can be considered an objective approach. There is room for other conventional priors.
\item{Regions of Statistical Importance:}
In order to state "statistical importance", rather than "statistical significance" the weight function can be combined with a loss structure to define "indifference regions" on which the difference between the null and alternative is of no practical importance. See the examples.
\end{enumerate}
     It  turns out that
under the prior measure interpretation, we obtain a Frequentist Decision Theory justification of Bayes Factors.
For the second interpretation, the decisions are based on posterior probabilities of sets, that actually embody "statistical importance" based rules.\\

Note: It is tempting, because it is so simple, to use weight functions which are point masses in the null and the point where statistical importance starts. These are point masses signalling specific points for which the errors ought to be controlled by design. For example if for a particular value of $\theta_1 \in \Theta_1$, where there is "practical significance", for example a novel medical treatment improvement of $20\%$, say, then the weight function may be set as a point mass on $20\%$ of improvement. (This typically would work only for monotone likelihood ratio families). These point masses rather than being though as sensible, we consider them for i) simplicity and ii) to compare it with frequentist solutions to the problem of "too much power", that is when there is statistical significance but not practical significance (Bickel and Doksum, 1977). However this simple solution, is not based on a reasonable prior (point masses).
\subsection{A General Optimality Result}
Define the weighted likelihoods, that we may call the \emph{\texttt{Evidences}}, for the $\mbox{Data}= \textbf{y}$
under each hypothesis as \be \varpi_0(\textbf{y})=\int_{\Theta_0}
f(\textbf{y}|\theta) \pi_0(\theta) d\theta, \ee and \be
\varpi_1(\textbf{y})=\int_{\Theta_1} f(\textbf{y}|\theta)
\pi_1(\theta) d\theta. \ee \textbf{Lemma 1:} It is desired to find a
test function $\delta$ that minimizes, for specified $a>0, \mbox{
and }, b>0$: \be \mbox { SERRORS}(\delta)=a\cdot \alpha(\delta) + b
\cdot \beta(\delta). \ee The test $\delta^*$ is defined as: accept
$H_0$, if \be \frac{\varpi_0(\textbf{y})}{\varpi_1(\textbf{y})}> \frac{b}{a}
, \ee and $H_1$ is accepted if \be \frac{\varpi_0(\textbf{y})}{\varpi_1(\textbf{y})}< \frac{b}{a}, \ee and accept
any if $a \cdot \varpi_0(\textbf{y})= b \cdot \varpi_1(\textbf{y})
$. Then for any other test function $\delta$: \be \mbox {
SERRORS}(\delta^*)= a\cdot \alpha(\delta^*) + b
\cdot \beta(\delta^*) \leq \mbox { SERRORS}(\delta),
 \ee
 in words rejecting the null when the ratio of evidences is smaller than $b/a$ is globally
 optimal.\\
 \verb"Proof":
Denote by \textrm{R}, the rejection region of the test $\delta$,
that is those data points on which $H_0$ is rejected. Then, under
the mild assumptions of Fubbini's Theorem that allows interchanging the order of the integrals,
for any test function $\delta$,
\[
a \alpha(\delta)+b \beta(\delta)=a
\int_{\Theta_0}[\int_\textrm{R}f(\textbf{y}|\theta)d\textbf{y}]\pi_0(\theta)
d\theta+b \int_{\Theta_1}[\int_{\textrm{R}^C} f(\textbf{y}|\theta)
d\textbf{y}]\pi_1(\theta) d\theta=
\]
\[
a  \int_{\Theta_0}\int_\textrm{R}f(\textbf{y}|\theta)
\pi_0(\theta) d\textbf{y} d\theta+ b
\int_{\Theta_1}\int_{\textrm{R}^C} f(\textbf{y}|\theta)
\pi_1(\theta) d\textbf{y}d\theta=
\]
\[
a  \int_{\Theta_0}\int_\textrm{R}f(\textbf{y}|\theta) \pi_0(\theta) d\textbf{y}d\theta+
b [1- \int_{\Theta_1}\int_\textrm{R}f(\textbf{y}|\theta) \pi_1(\theta)  d\textbf{y}d\theta]=
\]
\[
b+\int_\textrm{R} [a \int_{\Theta_0}
f(\textbf{y}|\theta)\pi_0(\theta)
d\theta-b\int_{\Theta_1}f(\textbf{y}|\theta)\pi_1(\theta)
d\theta]d\textbf{y}=
\]
\[
b+\int_\textrm{R}[a \varpi_0(\textbf{y})-b \varpi_1(\textbf{y})]d\textbf{y}.
\]
The results follows from application of the definition of $\delta^*$ in expressions $16$ and $17$,
since every point on which $a\cdot \varpi_0(\textbf{y})-b\cdot \varpi_1(\textbf{y})<0$ is in $\textrm{R}$, but no point on which
$a\cdot \varpi_0(\textbf{y})-b\cdot \varpi_1(\textbf{y})>0$, is included. Therefore $\delta^*$, minimizes the last term in the sum and the first does not depend on the test.
So the result has been established.\\
Regarding the assessment of the constants $a\mbox{ and }b$, notice
that it suffices to specify its ratio $b/a= \textrm{r} $. This can
be made as: i) By finding the \emph{implicit a and b} of a carefully designed experiment as in Example 1, ii) Conventional table of ratio of evidences, like in
Jeffreys Scale-Table of evidences, (See Appendix 1) or iii) by a ratio of prior
probabilities of $H_0$ times the loss incurred by false rejection of
$H_0$ over the prior probability of $H_1$ times the loss incurred by
false acceptance of $H_0$, in symbols, calling $L_0$ the loss for false rejection of $H_1$ and $L_1$ the loss for false rejection of $H_0$:
\[
\textrm{r}=\frac{b}{a}=\frac{P(H_1)\cdot L_0}{P(H_0)
\cdot L_1}.
\]
 To see this, notice that the  risk function can be
written as $R(\theta,\delta)=L_1 \alpha_{\theta}(\delta) \mbox{ if }
\theta \in \Theta_0$, and $ R(\theta,\delta)=L_0
\beta_{\theta}(\delta) \mbox{ if } \theta \in \Theta_1$. Assuming
that a priori the probability of the Null Hypothesis is $P(H_0)$,
then the average (Bayesian) risk, taking expectations with respect to $(P(H_0),
\pi_0)$ and $((1-p(H_0)), \pi_1)$, we get the averaged risk \be
r(\delta)=P(H_0)\cdot L_1 \cdot \alpha(\delta) + (1-P(H_0))\cdot L_0
\cdot \beta(\delta), \ee and it is seen that the correspondence with
expression (15) is: $a \mapsto P(H_0) \cdot L_1$ and $b \mapsto
(1-P(H_0)) \cdot L_0$ (assuming that the loss is constant on each of the hypothesis).\\
The Rejection Region \texttt{R} in (17) takes two different shapes according to the interpretations 1 and 2.
\begin{itemize}
\item For interpretation 1, \texttt{R} is defined as
\be
\frac{\varpi_0(\textbf{y})}{\varpi_1(\textbf{y})}< \frac{b}{a},
\ee
that is the Null Hypothesis is rejected if the Bayes Factor of $H_0$ over $H_1$ is small enough.
\item For interpretation 2, \texttt{R} becomes,
\be
\frac{Pr(H_0 \cup H_0^*|\textbf{y})}{Pr(H_1|\textbf{y})}=\frac{Pr(H_1^C|\textbf{y})}{Pr(H_1|\textbf{y})}<\frac{b}{a},
\ee
where $H_0$ is the null hypothesis, $H_0^*$ the indifference region where it is not worth to abandon the null because the improvement is not enough, and $H_1$ the alternative of practical significance. This assumes that the loss of rejecting $H_0$ under $H_0$ and $H_0^*$ is the same, and that the loss for accepting $H_0$ both under $H_0$ and $H_0^*$ is zero.\\
We put forward that (21) is more reasonable than, rejecting the null when
\[
\frac{Pr(H_0|\textbf{y})}{Pr(H_1|\textbf{y})}<1/3,
\]
say, (an approach popular in medical statistics), since it may happen than $P(H_0|\textbf{y})=0.1\varepsilon$ and $P(H_1|\textbf{y})=0.9\varepsilon$, and $\varepsilon$ can be minute, like $\varepsilon=0.001$. If both posterior probabilities are minute, you should \texttt{not} abandon $H_0$ in favor of $H_1$.\\
Finally, for the simple and simplistic point masses at $\theta_0$ and $\theta_1$ the optimal rule becomes,
\be
\frac{f(\textbf{y}|\theta_0)}{f(\textbf{y}|\theta_1)}<b/a.
\ee
\end{itemize}
\subsection{Relaxing the Assumptions: Predictive Priors Matching}
In the proof of Lemma 1, it was assumed that the weights be proper, i.e. $\int_{\Theta_j} \pi_j(\theta) d\theta=1$. This may be seen as a too heavy an assumption. Fortunately, the assumption can be relaxed, at least for (non-proper) weights which are improper but \emph{predictively matched}, see for example Pericchi (2005) for a discussion of predictively matched priors. We give two illustrations of predictively matched priors.
 \begin{enumerate}
 \item Illustration 1: Let us consider the prior Jeffreys suggested for the Normal mean testing problem: $H_0:\mu=\mu_0 \mbox{ VS } H_1: \mu \neq \mu_0$ and the variances are unknown. Jeffreys priors for this problem are: \\ $\pi^J_0(\sigma_0)=\frac{1}{\sigma_0}$ and \\
$\pi^J_1(\mu,\sigma_1)=\frac{1}{\sigma_1}\cdot \frac{1}{\pi \sigma_1 (1+\mu^2/\sigma^2_1)}$. \\ Notice that the priors are not proper. We call a training sample of minimal size $x(l)$, a sub-sample of $\mathbf{x}$ such that both $\pi^J_0(\sigma_0|x(l))$ and $\pi^J_1(\mu,\sigma_1|x(l))$ are proper, that is, the priors integrate 1, but any sub-sample of $x(l)$ will not. In this illustration  case the minimal training size is one, that is $x(l)=x_l$. It turns out that Jeffreys priors are predictively matched (Pericchi, 2005), that is, for whatever $x_l$,
\[
\int f(x_l|\sigma_0)\pi^J_0(\sigma_0) d\sigma_0=\int f(x_l|\mu, \sigma_1) \pi^J_1(\mu,\sigma_1)d\mu d\sigma_1,
\]
or $m_0(x_l)=m_1(x_l)$.
\item Illustration 2: Suppose you wish to compare a Normal model with a Cauchy model both with location $\mu$ and location $\sigma$ unknown. For location-scale models the objective prior is usually chosen as $\pi(\mu,\sigma)=1/\sigma$. It turns out, see  Berger, Pericchi and Varshavsky (1998) that for any location-scale likelihood, the minimal training sample size is 2, $x(l)=(x_{l_1},x_{l_2})$ and it turns out that,
    \[
    \int \frac{1}{\sigma^3}\cdot f((x_{l_1}-\mu)/\sigma)f((x_{l_2}-\mu)/\sigma)d\mu d\sigma=\frac{1}{2|x_{l_2}-x_{l_1}|},
    \]
    that is the predictive for any two different data point, is the \emph{same} for \emph{any} location-scale family, and thus, if the prior is $1/\sigma$ any location-scale family is predictively matched to any other location-scale family for the minimal training sample of 2 observations.
\end{enumerate}
\textbf{Corollary 1} For priors which does not integrate 1, but are predictively matched, Lemma 1 still holds. \\
\textbf{Proof:} Take an arbitrary minimal training sample $x(l)$, so that the remaining sample is denoted y $x(-l)$. Now use the priors $\pi_0(\theta_0|x(l))$ and $\pi_1(\theta_1|x(l))$, and corresponding likelihoods $f_0(x(-l)|\theta_0)$ and $f_1(x(-l)|\theta_1)$ in Lemma 1. Assuming we have a sample bigger than the minimal training sample, then Lemma 1 follows with the priors and likelihoods above. Now the result follows from the following identity, for predictively matched priors:
\[
\frac{\int f_0(x(-l)|\theta_0) \pi_0(\theta_0|x(l))d\theta_0}{\int f_1(x(-l)|\theta_1) \pi_1(\theta_1|x(l))d\theta_1}=
\]
\[
\frac{\int f_0(x|\theta_0) \pi_0(\theta_0)d\theta_0}{\int f_1(x|\theta_1) \pi_1(\theta_1)d\theta_1}.
\]
This Corollary, enlarge substantially the applicability of Lemma 1 and highlights the fundamental importance of predictively matched priors.
\section{Two Sided Alternatives Example 2:}This is an univariate Normal Example with known variance $\sigma_0^2$ and the hypothesis center on the mean $\theta$:
\[
H_0: \theta=\theta_0 \mbox{ VS } H_1: \theta \neq \theta_0.
\]
\subsection{Bayesian Intrinsic Prior:} One objective Bayes approach is the Intrinsic Prior approach. For this hypothesis calculation it turns out that the Intrinsic Prior is (Berger and Pericchi(1996) and Pericchi (2005)): $\pi_1(\theta)=N(\theta|\theta_0, 2 \sigma^2_0)$ and $\pi_0$ is a Dirac Delta at point $\theta_0$. Calculation yields that the optimal test $\delta^*$ is: Reject $H_0$ if
\be
\frac{f(\bar{y}|\theta_0)}{\varpi_1(\bar{y})}=\frac{N(\bar{y}|\theta_0, \sigma_0^2/n)}{N(\bar{y}|\theta_0, \sigma_0^2(2+1/n))}< \textrm{r},
\ee
where $n$ is the sample size.
\subsection{Practical Significance:} Assume that the test is aimed to detect a difference if $\theta_1=\theta_0 \pm \Delta$. The the simplistic prior weight is a Dirac Delta centered at the two points $\theta_0\pm \Delta$ with weight equal to $1/2$ on each point. The test now becomes: Reject $H_0$ if
\be \frac{f(\bar{y}|\theta_0)}{\varpi^*_1({\bar{y}})}=
\frac{N(\bar{y}|\theta_0,\sigma^2_0/n)}{\frac{1}{2}
N(\bar{y}|\theta_0-\Delta,\sigma_0^2/n)+\frac{1}{2}N(\bar{y}|\theta_0+\Delta,\sigma_0^2/n)}<
\textrm{r}, \ee which can be written as
\[
\exp(-\frac{n\Delta}{2\sigma_0^2}[\Delta-2(\bar{y}-\theta_0)])+
\exp(-\frac{n\Delta}{2\sigma_0^2}[\Delta+2(\bar{y}-\theta_0)])>\frac{2}{r},
\]
a reasonable criterion, that can be compared with the usual significance test which is extremely biased against $H_0$ for sample sizes
which are larger than the carefully designed sample sizes to achieve a specified Type I and Type II errors.\\
A more careful analysis of indifference regions, using the same Intrinsic prior as above lead us to the following rejection region:
\be
\frac{Pr(H_{0,\Delta}|\textbf{y})}{1-Pr(H_{0\Delta}|\textbf{y})}<b/a,
\ee
where $H_{0,\Delta}=[\theta_0-\Delta,\theta_0+\Delta]$.  The resulting criteria is extremely simple, based in the
ratio of two Normal Probabilities, that takes specific account of the indifference region.
It can be verified that all tests (23), (24) and (25), are consistent as the sample size
$n$ grows. That is \textbf{both} type I and type II errors go to zero as the sample size grows.\\
Now, the alternative paradigm enjoys several desirable properties, some of which we proceed to describe.
\section{Now Hypothesis Testing under the new paradigm obeys the Likelihood Principle.}
One of the usual criticisms against significance testing is that it does not obey the Likelihood Principle, a Principle not only shared by Bayesians, but that was actually enunciated or defended by eminent non-Bayesians. Loosely speaking the Likelihood Principle establishes that
if two likelihoods are proportional to each other, the information about
the parameter vector $\theta$ is the same. The following important example is eloquent.
\subsection{Example 3: Lindley and Phillips (1976) Example revisited:}
It is desired to test if a coin is balanced, since it is suspected that it is more prone to Heads:
\[
H_0: \theta=1/2, \mbox{ VS } H_1: \theta>1/2.
\]
It is known that the number of Heads is $S=9$ and the number of tails is
$n-S=3$. It is desired to conduct a test at $\alpha=0.05$, shall we reject the Null. The astonishing fact is significance testing (in its two current versions, based on p-values or fixed significance) can not decide. How the sample size was decided? It was fixed beforehand? Or it was decided that the experiment would stop at the third tail? Or the statistician stopped when her door was knocked by a student? The answers are not the same!\\
Suppose that $n=12$ was decided beforehand, then we have a binomial likelihood,
\be
f_B(S|\theta) = \frac{12!}{9!3!} \theta^9 (1-\theta)^3=220 \theta^9 (1-\theta)^3,
\ee
but if the experiment was stopped at the third tail, we have a Negative Binomial experiment,
\be
f_{NB}(S|\theta)=\frac{11!}{9!2!} \theta^9 (1-\theta)^3= 55 \theta^9 (1-\theta)^3,
\ee
that is we have two proportional likelihood functions, and according to the Likelihood Principle we should have the same inference. However,
the observed p-values differ:
\[
\alpha_B=Pr(S\ge 9|\theta=0.5,\mbox{Binom})=\sum_{S=9}^{12}f_B(S|\theta=0.5)=
0.073,
\]
while
\[
\alpha_{NB}=Pr(S\ge 9|\theta=0.5,\mbox{NegBinom})=\sum_{S=9}^{\infty}f_{NB}(S|\theta=0.5)=
0.0327,
\]
thus the results are statistically significant in the second scenario but not in the first.
We have not attempted to model the third situation that someone knocks at your door making you to stop, but the result will surely be still different, with the same outcome! \\
Examples like these seem to have convinced many that frequentist Hypothesis Testing is \textbf{bound} to violate the Likelihood Principle. The good news, surprising to many we would guess,  is that the violation of the Likelihood Principle is avoided by DeGroot's method, that is if the weighted sum of type I and type II errors is minimized.\\
\textbf{Corollary 1.} Testing by Minimizing a weighted sum of errors, automatically obeys the Likelihood Principle.\\
\texttt{Proof:} From Lemma 1, the optimal test is Reject $H_0$ if
\[
\frac{\varpi_0(\textbf{y})}{\varpi_1(\textbf{y})}=\frac{\int_{\Theta_0}
f(\textbf{y}|\theta) \pi_0(\theta) d\theta}{\int_{\Theta_1}
f(\textbf{y}|\theta) \pi_1(\theta) d\theta}<b/a,
\]
and in the ratio of the left hand side, the constant in the likelihood cancel out.\\
\textbf{Example 3 (Continued)} In this example we are going to assume the Uniform Prior in $(0.5,1)$, $\pi(\theta)=2\times 1_{(0.5,1)}(\theta)$. We assume this prior for simplicity although we do not think it is "optimal" in any sense, it is not unreasonable and does not influence the outcome heavily. Then the ratio of evidences is easily found numerically,
\[
\frac{f(S|\theta=0.5)}{\int^1_{0.5}f(S|\theta)\cdot 2 d\theta}=
\]
\[
\frac{(1/2)^{12}}{(1-pbeta(0.5|10,4)\times Beta(10,4)\times 2)}=0.366,
\]
where $pbeta(x|a,b)$ is the probability of a beta density, parameters a and b, from zero to $x$, and Beta is the Beta function $Beta(a,b)=\frac{\Gamma(a)\cdot \Gamma(b)}{\Gamma(a+b)}, a>0, b>0$.\\
Thus according to Jeffreys table of evidence, the ratio is smaller than 1 but bigger that $1/\sqrt{10}=0.32$, so there is mild evidence against $H_0$.\\
The fact that traditional Hypothesis Testing depends, in unnatural ways, to the stopping rule is one of its weakest links. It has far reaching consequences for example in clinical trials, where quite often the trials has to be cut short for example for ethical reasons. Procedures that depend on ratios of probabilities rather than tail probabilities are more realistic more flexible and more ethical.
\section{When Statistical Significance meets Practical Significance}
One of the most criticized points of the current significance testing approach is the lack of correspondence between practical significance and statistical significance. One such example is in Freeman (1993).
\subsection{Example 4: Freeman's Example }
Consider four hypothetical studies in which equal number of patients are
given treatments A and B and are asked which of them they prefer. The results are in the following table.\\

\begin{tabular}{|l|l|l|l|l|}
  \hline
 Number of patients  & Number of patient  & Percentage  &two-sided  & Ratio of \\
 receiving A and B & preferring A:B & preferring A & P-value & Evidences \\
 \hline
  20 & 15:5 & 75.00 & 0.04 & 0.42\\
  200 & 115:86 & 57.50 & 0.04 & 1.85\\
  2000 & 1046:954 & 52.30 & 0.04 & 6.75\\
  2 000 000 & 1 001 445 : 998 555 & 50.07 & 0.04 & 219.66 \\
  \hline
\end{tabular}

As a weight function for the parameter $\theta$ an objective(and proper)prior is the Jeffreys' prior,
\[
\pi^J(\theta)=\frac{1}{\pi} \theta^{-1/2} (1-\theta)^{-1/2},\mbox{  for  } 0<\theta<1.
\]
Computation yields, that the ratio of evidences is
\[
\frac{f(s|\theta=1/2)}{\varpi(s)}=\frac{\pi \cdot 0.5^N}{Beta(s+1/2,N-s+1/2)},
\]
which leads, see the 5th column in the table consistent with the conclusions put forward by Freeman on intuitive
grounds:  the first trial is too small to permit reliable conclusions while the last trial would be considered as
evidence \emph{for}  rather than against equivalence, since for any practical perspective the two treatments are equivalent. In fact the ratio gives "decisive" evidence or grade 5 in favor of the Null Hypothesis for 2 Million patients.
\section{Is there Extra Sensorial Perception ESP? Or Just Humongous Large Numbers?}
In one of their books, Wonnacott and Wonnacott declared: "Do you want to reject a hypothesis? Take a large enough sample!"
\subsection{Example 5: ESP Example:}
The so called Extra-sensorial Experiment, Good (1992) is a good example how the p-values are increasingly misleading with extremely large samples. \\
\textbf{Example 5:}{\textbf{Extra Sensorial Perception: ESP or no ESP?} Here the question is
if a "gifted" individual can change the proportion of 0's and 1's, that are emitted with "perfectly" balanced proportions. So the Null is no change
in the proportion against some change or:  $H_0:\theta=1/2 \mbox{ VS } H_1: \theta \neq 1/2$.
We have a Huge sample: $N=104,490,000$;
Successes: $S=52,263,471$;
Ratio: $S/N=0.5001768$;

The p-value against the null is minute: $pval=0.0003$, leading to a compelling rejection of $H_0$.\\
On the other hand an objective (proper) prior exists here, which is the Jeffrey's prior, that may be used as a weight function:
 (Jeffreys') Prior: $\pi^J(\theta)=\frac{1}{\pi \times \sqrt{\theta (1-\theta)}}$. Then the Bayes Factor, or the \textbf{ratio of evidences}, is
\[ B_{H_0,H_1}=\frac{f(data|\theta=1/2)}{\int f(data|\theta) \pi^J(\theta) d\theta}=\frac{\pi \cdot (1/2)^N}{Be(S+0.5,N-S+0.5)}= \]
\[B_{H_0,H_1}=\exp(2.93)=18.7, \]
which is a strong support of the null hypothesis. (The Bayes factor  is equal to 12 for the Uniform prior, also a strong support of the Null) \\
Taking the second route, of setting the apriori points of \textbf{practical significance}, we may agree that say
$\Delta$, above or below the null can
be acceptable as practical significance. In general call $\Delta$ the percent accepted as practical. The criterion now read as:
\be
Pr(0.5-\Delta<\theta<0.5+\Delta|data)/(1-Pr(0.5-\Delta<\theta<0.5+\Delta|data))<r=b/a.
\ee
\begin{table} \begin{center}
\begin{tabular}{|l|r|r|r|r|}
  \hline
  $\Delta$ & 0.0002 & 0.0003 & 0.0004 &  0.0005\\
  r        & 2.15 & 169 & 397877 & 51369319698\\
  \hline
\end{tabular}\caption{Table for different values of $\Delta$}
\end{center}
\end{table}

We run the $\Delta$ from $0.0002$ (or $0.02\%$) to $0.0005$ or $0.05\%$. In Table 1 even for the smallest $\Delta$ the ratio of
likelihoods is bigger than one and for $\Delta=0.0005$ it grows to compelling evidence against $H_1$. This is in sharp contrast with a p-value of $0.0003$.
\section{A General Inequality: the discrepancy between fixed significance levels and minimizing sum of errors is general}
Even though the discrepancies between fixed significance levels and minimizing sum of errors have been illustrated through examples, it is a general phenomenon,
as shown in the following result (See also, Birnbaum (1969) and Dempster (1997), for related results).\\
\textbf{Lemma 2:} For the optimal test $\delta^*$ of Lemma 1, it turns out that:
\[
 \frac{\alpha(\delta^*)}{1-\beta(\delta^*)}\leq \frac{b}{a}.
\]
\textbf{Proof:} First notice that the rejection region for the test $\delta^*$ can be written as: $\texttt{R}=\frac{\varpi_1(y)\cdot b}{\varpi_0(y) \cdot a}\ge 1$. Call the set $\texttt{S} \subset \texttt{R}$ where $\varpi_0(y) > 0$ Now,\\
\[
\alpha(\delta^*)=\int_\texttt{R}\int_{\Theta_0} f_0(y|\theta)  \pi_0(\theta) d\theta dy=\int_\texttt{S} \varpi_0(y) dy=
\]
\[
\leq \int_\texttt{S}\frac{\varpi_1(y) b}{\varpi_0(y) a} \varpi_0(y)dy=
\]
\[
\frac{b}{a}\int_{\texttt{S}} \varpi_1(y) dy \leq \frac{b}{a} \int_{\texttt{R}} \varpi_1(y) dy=\frac{b}{a}(1-\beta(\delta^*)).
\]
\textbf{Corollary 2}:
\[
\alpha(\delta*) \leq \frac{b}{a}.
\]
Thus for example, if $b/a=20$ then $\varpi_1(y^*)/\varpi_0(y^*)$ is considered equivalent to $\alpha(\delta^*)=0.05$ and the power is $0.8$ then
\[
\frac{\varpi_1(y^*)}{\varpi_0(y^*)}\leq \frac{1-\beta}{\alpha}=\frac{.8}{.05}=16,
\]
and the Corollary 2 that $\frac{\varpi_1(y^*)}{\varpi_0(y^*)}\leq 20$,
so DeGroot's test rejects less often.

\section{Conclusions} What emerges in the implementation and extensions of DeGroot's approach to Testing Statistical Hypothesis, is a practical implementation of the ideas of Decision Theory with a bridge between Bayesian and frequentist philosophies. This implementation, we argue along with DeGroot, is superior to the two implementations dominant in practice: i) The use of p-values with fixed cut points, like the ubiquitous: $\mbox{ $\alpha$-set}[0.1*, 0.05**, 0.01***]$, and ii) the use of fixed Type I errors in the $\alpha$-set, and then choosing the criterion to minimize Type II error.\\
By doing i) or ii) you are in danger of having a minute effective Type II error relative to an enormous Type I error. Furthermore, by fixing Type I error, inconsistency is obtained "by design": \textbf{no matter how informative the experiment is, you force the method to have no-less than a Type I error in the $\alpha$-set}. On the contrary, minimizing the weighted sum or errors, a more balanced (between two errors) method emerges and consistency flows as a automatic consequence: \textbf{By minimizing the sum of errors as evidence grows you are letting both errors converge to zero so you have a consistent method}. As virtues of the approach we have a general theory of optimal testing, that obeys the Likelihood Principle and the Optional Stopping Principle, reconciles the disagreement between schools of statistics and is more in line with the demands of the scientific method.\\
Finally, to achieve the benefits of the general theory, is not necessary to assume fully proper priors: predictively matched improper priors suffices.
\section{References}
\begin{enumerate}
\item Bayarri, M.J. and Berger, J.O. (2004) The interplay between Bayesian and frequentist analysis.  `Statistical Science', {19}, 58--80.
\item Berger, J.O. (2008) A comparison of testing methodologies. In the proceedings of the PHYSTAT-LHC Workshop on Statistical Issues for LHC Physics, June 2007, CERN 2008-001, pp. 8-19.

\item Berger J.O. Pericchi L.R. and Varshavsky J.A (1998) Bayes Factors and Marginal
Distributions in Invariant Situations. Sankya: The Indian Journal of Statistics.
Special Issue on Bayesian Analysis. Vol 60, Series A, part.3 , p.307-321
\item Bickel P.J. and Doksum K.A. (1977) Mathematical Statistics: Basic Ideas and Selected Topics. Holden-Day, Inc.

\item Birnbaum, A. (1969) Concepts of statistical evidence. In \emph{Philosophy, Science and Methods: Essays in Honor of Ernest Nagel.} (eds
S. Morgenbesser, P. Suppes and M. White) St. Martin's Press, New York
\item Cox, D.R. and Hinkley, D.V. (1974) Theoretical Statistics. Chapman and Hall.

\item DeGroot, M. (1975) \textit{Probability and Statistics}, 2nd Edition. New York. Addison-Wesley.
\item Dempster, A.P. (1997) The direct use of likelihood for significance testing. \textbf{Statistics and Computing}, 7, p. 247-252.
\item Ioannidis, J. P. A. (2005) Why most published research findings are false. PLoS Medicine August vol2 issue 8 e24, p. 696-701

\item Lindley, D.V. (1957) A statistical paradox. Biometrika, 44,p. 187-
192.

\item Good, I.J. (1992) The Bayes/Non-Bayes Compromise: A Brief Review. Journal of the American Statistical Association, Vol. 87, No. 419, pp. 597-606.

\item Lindley, D.V. and Phillips, L.D. (1976) Inference for a Bernoulli
process (a Bayesian view). Amer. Statist.,30,p. 112-129.
\item Pereira, CAB (1985), Testing hypotheses of diferent dimensions: Bayesian View and Classical Interpretation (in Portuguese), Professor Thesis: the Institute of Mathematics and Statistics. Universidade de S\^{a}o Paulo.

\item Pereira CAB and Wechsler S (1993), On the concept of P-Value, Brazilian J Probability and
Statistics 7;159-77.
\item Perez, M.E. and Pericchi, L.R. (2012) Changing Statistical Significance as the Amount of Information Changes: The Adaptive $\alpha$ Significance Level. Technical Report Biostatistics and Bioinformatics Center, UPR-RRP.
\item Pericchi, L.R. (2005) Model Selection and Hypothesis Testing based on Objective Probabilities and Bayes Factors. \emph{Handbook of Statistics Vol. 25}, Dey D.K and Rao C.R. Editors. pages 115-149.
\item Siegfried, T. (2010) Odds Are, It's Wrong
Science fails to face the shortcomings of statistics. ScienceNews,  March 27th, 2010; Vol.177 \#7 (p. 26)

\item Varuzza L; Pereira CAB (2010), Significance test for comparing digital gene expression profiles: Partial likelihood application, Chilean J Statistics 1(1):91-102.
\end{enumerate}

\section{Appendix 1: Jeffreys Table of Evidences}
Here $r=b/a$
\begin{tabular}{|l|r|r|}
  \hline
  Grade 0& $r \ge 1$& Null Supported \\
  Grade 1 &$ 1>r>10^{-1/2}$& Mild Evidence against $H_0$ \\
  Grade 2 & $10^{-1/2}>r>10^{-1}$ & Substantial Evidence against $H_0$ \\
  Grade 4 & $10^{-1}>r>10^{-3/2}$ & Strong Evidence against $H_0$ \\
  Grade 5& $10^{-3/2}>r>10^{-2}$ & Very Strong Evidence against $H_0$ \\
  Grade 6& $10^{-2}>r$ & Decisive Evidence against $H_0$ \\
  \hline
\end{tabular}
\end{document}